\documentstyle[preprint,eqsecnum,aps]{revtex}
\begin{document}
\draft
\tightenlines
\preprint{MSUCL-1111}
\title{Effects of Isospin Asymmetry and In-Medium Corrections 
       on Balance Energy}
\author{Frank Daffin and Wolfgang Bauer\cite{hisemail}}
\address{
National Superconducting Cyclotron Laboratory, Michigan
State University\\
East Lansing, Michigan 48824--1321}

\maketitle 
\begin{abstract}
The effects of an isospin asymmetry and in-medium corrections to the nucleon
collision cross section on the balance energy are explored.  The BUU model for
intermediate energy heavy-ion collisions is used with isospin-dependent
mean fields to calculate the balance energies of $^{58}$Fe + $^{58}$Fe
and $^{58}$Ni + $^{58}$Ni for a range of impact parameters.  We find that
we are able to reproduce the impact parameter dependence of the balance
energy, and the sign (but not the magnitude) of the shift in balance 
balance energy as a function of isospin asymmetry.
\end{abstract}

\narrowtext

\section{Introduction}
\label{sec:intro}
The bulk properties of nuclear matter under extreme conditions are of interest
to both nuclear physics and astrophysics.  To generate these conditions,
we collide nuclei in search for hints
on the nature of the nucleus-nucleus interaction.  Theoretical studies of
these events allow us to make a connection between models and
experiment though observables such as the balance
energy \cite{kr89,ogi89,west93}.  Balance energy, or the beam energy at which
transverse, in-plane directed flow \cite{po85} 
changes sign as a function
of beam energy, is sensitive to the forces underlying the dynamics of these
collisions.  Thus its study can yield valuable information about the complex
interplay of the repulsive nature of the hard scattering nucleon cross section,
the repulsion and attraction of the nuclear mean field, and the repulsive
contribution of the Coulomb force.

The BUU (Boltzmann-Uehling-Uhlenbeck) 
\cite{bk84,kjm85,ab85,bbc86,sg86,bg88} transport model is 
successful in reproducing
and predicting experimental magnitudes and trends in collective single-particle
observables \cite{bl87,gw90,wo90}, such as the balance energy, and we use it
here. BUU is a theory for the single-particle phase space distribution of
baryons (plus resonances and mesons).  In its most popular numerical 
realization, 
the model propagates test particles with Hamilton's equations through the
influence of experimentally observed nucleon collision cross
sections \cite{pdg88}, the Coulomb field, and a nuclear mean field based on
a Skyrme-type interaction \cite{sk59} which typically 
depends upon local density and momentum.

However, the inter-nucleon potential is not symmetric
with respect to the isospin of the nucleon.  Protons and neutrons experience
different forces due to their constituents.  This means that an asymmetry
term exists in the nuclear mean-field.  In addition, the nucleon cross
section is modified from its vacuum values in nuclear
matter \cite{bru55,kla93,alm95}.

Work by M\"{u}ller and Serot \cite{mul95} has shown that the liquid-gas phase
transition may in fact be of second rather than first order.  Their
work, based upon a thermodynamic approach, demonstrated that local chemical
instabilities are responsible for the transition when an asymmetric potential
is used.  This contrasts with the common view that mechanical instabilities are
responsible for the onset of fragmentation in excited, diffuse systems.
Studies using the percolation model \cite{bau84}
also have been used to explore the
isospin degree of freedom in the nuclear fragmentation 
transition \cite{kor97,bau98}.
The new experimental facilities coming on-line which will probe heavy-ion
dynamics near the drip lines affords an opportunity to explore these results.

In stellar evolution the softening of the compressibility of nuclear matter as
neutron ratios deviate from 1/2 is critical for generating a supernova
explosion for massive stars (12\,M$_{\odot}$ to 15\,M$_{\odot}$) \cite{bar85}.
Core collapse cannot sustain a shock if the nuclear Equation
of State (EoS) retains its equilibrium
stiffness; the result is at least a delay of such important process as
nucleosynthesis.

Neutron stars cool by emitting neutrinos.  After an initial fast-cooling phase,
they enter a more sedate cooling epoch producing neutrinos via the 
modified URCA process: $(n,p)+p+e^{-}\rightarrow(n,p)+n+\nu_{e},\ \
(n,p)+n\rightarrow(n,p)+p+e^{-}+\bar{\nu}_{e}$.
However, if the proton density exceeds a critical value between 11\% and
15\%, which is dependent upon the symmetry energy, the direct URCA process
can occur and become the dominant cooling mechanism\cite{lat91}:
$n\rightarrow p+e^{-}+\bar{\nu}_{e},\ \
p+e^{-}\rightarrow n+\nu_{e}$.

The exploration of the isospin degree of freedom in heavy ion collision
thus carries the promise that it will help determine the answers to these
questions of astrophysical relevance \cite{lkb98}.

Recently, it was experimentally shown \cite{pak97} that the balance energy
for symmetric systems of equal mass has different values for different
isospin asymmetry of the colliding nuclei.  In this paper we therefore
study the isospin-dependence of the balance energy, attempt to isolate
its origin, and hope to find sensitivity to the isospin-dependence of the
nuclear mean field.

\section{Model Input}

The BUU transport equation evolves in time a single nucleon phase-space
distribution under the influences of nucleon-nucleon collisions, the Coulomb
field, and a nuclear mean field. In this study, we choose two mean fields:
One recently used by Bao-An Li, {\it et al.} \cite{li95,li96} (also see 
Ref.\,\cite{tsa89,dan92,li93}):
\begin{equation}
      U = A \left( \frac{\rho}{\rho_0} \right) +
          B \left( \frac{\rho}{\rho_0} \right) ^{\sigma} +
          C \tau_z \left( \frac{\rho_n-\rho_p}{\rho_0} \right),
      \label{equation:limeanfield}
\end{equation}
where $\rho_0$ is the normal nuclear density, $\rho_n$ is the neutron density,
$\rho_p$ is the proton density, and $\tau_z$ is the isospin factor which is
$1$ for neutrons and $-1$ for protons.  Coefficients $A$, $B$ and $\sigma$ are
typically chosen to match the ground state properties of symmetric nuclear
matter such as the saturation density and saturation binding energy.  The
compressibility is a free parameter, and in this study we chose it to be
$200$~MeV.  This gives $A=-109$~MeV, $B=82$~MeV, and $\sigma=\frac{7}{6}$.  In
keeping with Bao-An Li's work, $C=32$~MeV.

The other mean field is derived from a Hamiltonian due to Sobotka \cite{sob94}:
\begin{equation}
      {\cal H} = {\cal KE} + \frac{4a}{\rho_0}({\rho_n}^{2}+
                 b\rho_n\rho_p+{\rho_p}^2)+
                 \frac{4c}{\rho_0}({\rho_0}^2\rho_p+\rho_n{\rho_p}^2).
      \label{equation:sobhamil}
\end{equation}
Here the coefficients $a=-3.66$~MeV, $b=15.0$ and $c=23.4$~MeV and ${\it KE}$
is the kinetic energy term.  
Neutrons are acted upon by:
\begin{equation}
      U_n = \frac{\partial{\cal H}}{\partial{\rho_n}} = 
      8a\frac{\rho_n}{\rho_0} + 4ab\frac{\rho_p}{\rho_0} +
      8c\frac{\rho_n\rho_p}{{\rho_0}^2} + 4c\frac{{\rho_p}^2}{{\rho_0}^2},
      \label{equation:sobmean}
\end{equation}
whereas protons are affected by:
\begin{equation}
      U_p = \frac{\partial{\cal H}}{\partial{\rho_p}} =
      8a\frac{\rho_p}{\rho_0} + 4ab\frac{\rho_n}{\rho_0} +
      8c\frac{\rho_n\rho_p}{{\rho_o}^2} + 4c\frac{{\rho_n}^2}{{\rho_0}^2}.
\end{equation}
In symmetric matter, this mean field reduces to the often-called ``Stiff''
equation of state (compressibility $380$~MeV).  It should be noted 
that there is
no transformation of the mean field of Sobotka that will yield an equivalent
``Soft'' equation of state.  Thus direct comparisons between the two
formulations are impossible.  One may simply alter $A$, $B$ and $\sigma$ in
Equation\,\ref{equation:limeanfield} to give the appropriate compressibility in
symmetric matter.

\begin{figure}[htb]

\vspace*{6cm}

\includegraphics{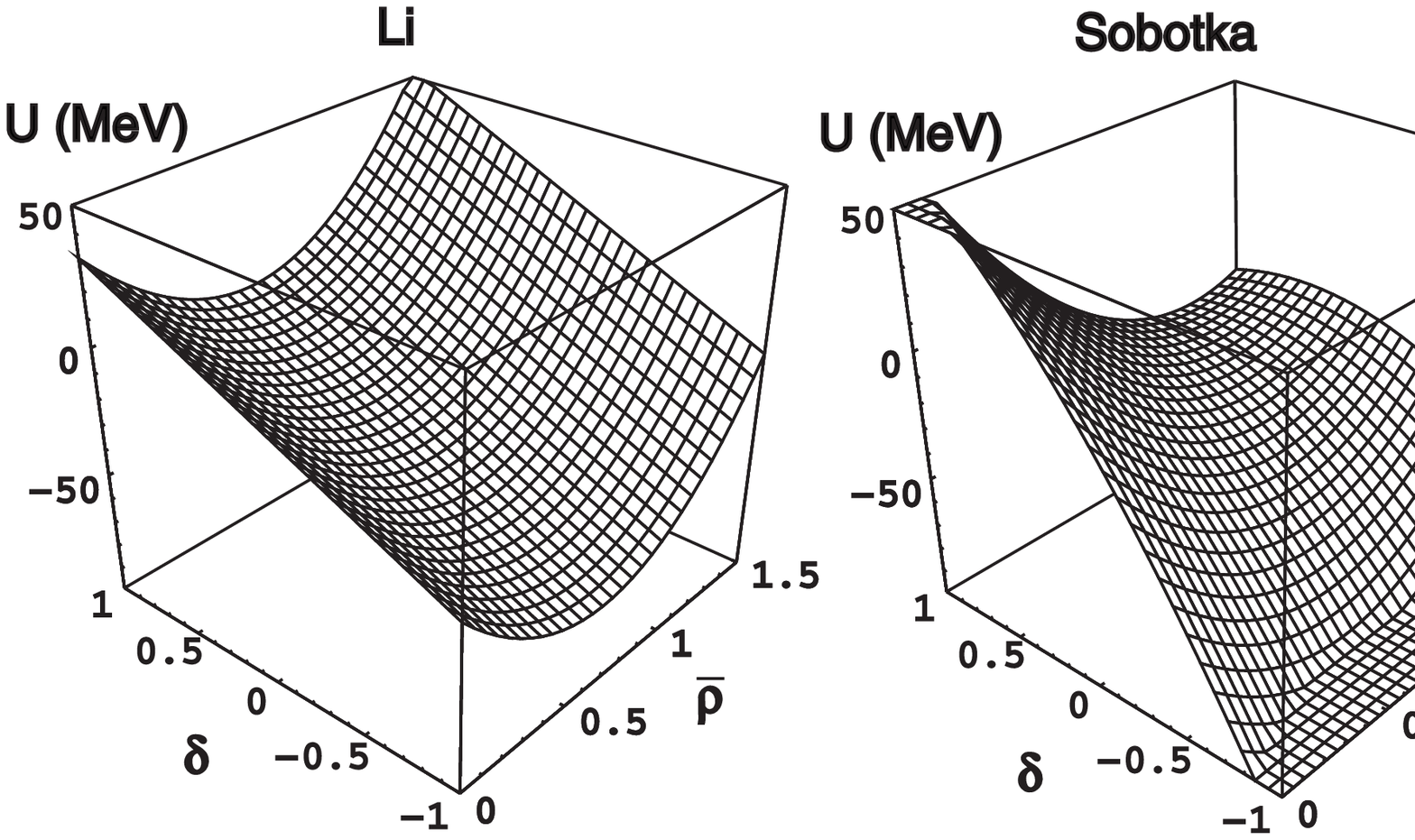}

\vspace*{7cm}

\caption[]{Mean field potential as a function of reduced density,
$\hat\rho = \rho/\rho_0$, and isospin asymmetry, $\delta$, as used
by Li et al.\ (left) \cite{li95,li96} and Sobotka (right) \cite{sob94}.}
\label{fig:one}
\end{figure}

With the isospin asymmetry
\begin{equation}
   \delta=\frac{\rho_{n}-\rho_{p}}{\rho_{0}}
   \label{equation:delta}
\end{equation}
and the reduced density
\begin{equation}
  \overline{\rho}=\frac{\rho}{\rho_{0}}
  \label{equation:newrho}
\end{equation}
Equations\,\ref{equation:limeanfield} and \ref{equation:sobmean} can be
written as:
\begin{equation}
   U = A\overline{\rho}+B\overline{\rho}^{\sigma}+C\tau_{z}\delta
   \label{equation:lidelta}
\end{equation}
and
\begin{equation}
     U_{n}=\overline{\rho}(4a+2ab)+\delta(4a-2ab)-c\delta^{2} + 
           3c\overline{\rho}^{2}-2c\delta\overline{\rho},
     \label{equation:sobdelta}
\end{equation}
for the neutron potential, and
\begin{equation}
      U_{p}=\overline{\rho}(4a+2ab)+\delta(2ab-4a)-c\delta^{2}
            +3c\overline{\rho}^{2}+2c\delta\overline{\rho},
\end{equation}
where $a$, $b$ and $c$ are as defined above.  Figure\,\ref{fig:one} (left side)
shows Equation\,\ref{equation:lidelta} as neutron excess and normalized nuclear
density are varied.  The right side of the same figure shows 
Equation\,\ref{equation:sobdelta} as neutron excess and normalized nuclear
density are varied.  One can see that
Equation\,\ref{equation:sobdelta} is more attractive to neutrons in the midst
of proton-rich matter for all densities than Equation\,\ref{equation:lidelta}.

Other parameterization of nuclear mean field potential also exist, \cite{ms98}
but were not considered here.

The nucleon-nucleon cross sections are parameterization from the Particle Data
Group\cite{pdg88} with medium modification implemented according to the
density dependent prescription:
\begin{equation}
{\boldmath \sigma}_{\rm NN} = {\boldmath \sigma}^{\rm free}_{\rm NN}\,(1
                               +\alpha~\overline{\rho})
\label{equation:corr}
\end{equation}
where $\alpha$ is the logarithmic derivative of the in-medium cross section
with respect to the density, taken at $\rho=0$,
\begin{equation}
      \alpha = \rho_0\,\frac{\partial}{\partial\rho}\left.\left(\ln
      {\boldmath \sigma}_{\rm NN}\right)\right|_{\rho=0}
\end{equation}
It is the lowest
coefficient of a Taylor-expansion of the cross-section in powers of the
density \cite{kla93}.
This is a parameterization of the Pauli-blocking of intermediate states is
motivated by Br\"{u}ckner $G$-matrix theory \cite{bru55}.  One can show
\cite{alm95,schulze97,schnell98} that values of $\alpha$ in the range between
$-0.4$ and $-0.2$ yield the best agreement of the simple parameterization
used here and the more involved $G$-matrix calculation using a realistic
nucleon-nucleon interaction.
In addition, it has been shown that an in-medium reduction of 
$\alpha\approx -0.2$ provides the best reproduction of 
the flow signals of symmetric collisions for a broad range of
energies and projectile-target combinations \cite{ogi90,kla93,west93}.

\section{Elementary considerations}

Before we proceed to show our numerical results, it is appropriate to
attempt to understand the reasons why we would expect a difference in
the balance energy for symmetric systems of equal mass, but different
isospin.
There are three isospin effects that we may consider: the average effective
nucleon-nucleon cross section, the difference in the Coulomb potential,
and the isospin dependence of the mean field.  

In order to understand the sign of the different effects, we first remind 
ourselves that the balance energy marks the beam energy at which the
repulsive and attractive interactions are roughly equal.  At beam energies
lower than the balance energy, the attractive interaction dominates, and
at energies higher than the balance energy, repulsive interactions determine
the flow.

The difference in the Coulomb interaction is very simple.  Fe has 26 protons
and Ni 28.  This means that the ratio of the Coulomb forces is $(26/28)^2$ =
0.862. Since the Coulomb force adds repulsion, we thus expect that the balance
energy for nickel should be lower than for iron.

The difference in the isospin dependent mean field has the opposite effect.
The isospin asymmetry, $\delta$, is three times
bigger for iron than it is for nickel:
\begin{equation}
  \delta(^{58}Fe) = 0.103,\ \ \ \ \delta(^{58}Ni) = 0.034
\end{equation}
A larger isospin asymmetry results in more
repulsion, compare Fig.\ 1. Thus we expect that the 
isospin dependent mean field alone would favor a lower balance energy
for the $^{58}$Fe system than for the $^{58}$Ni system.

The numbers of protons and neutrons in the colliding system also have an
influence on the average nucleon-nucleon cross section, because the 
neutron-proton cross section is approximately a factor of 3 bigger than the
proton-proton cross section in the energy regime of interest here.  We can
then calculate the average nucleon-nucleon cross section from the number
of collisions between nucleons of equal and opposite isospin,
\begin{eqnarray}
   \overline{\sigma}
   &=& \frac{N_{np}\,\sigma_{np} + (N_{nn} + N_{pp})\,\sigma_{pp}}
            {N_{np}              + N_{nn} + N_{pp}}\nonumber\\
   &=& \frac{3\,N_{np} + N_{nn} + N_{pp}}
            {   N_{np} + N_{nn} + N_{pp}}\,\sigma_{pp}
\end{eqnarray}
where $N_{np}$ is the number of neutron-proton collisions.

We make the {\em Ansatz} 
that the number of collisions of two types of
nucleons is proportional to the number of nucleons of each species, taken
to some power, $\zeta$,
\begin{eqnarray}
   N_{np} &\propto&(N_{n,T}\,N_{p,P})^\zeta+(N_{n,P}\,N_{p,T})^\zeta\nonumber\\
   N_{nn} &\propto&(N_{n,T}\,N_{n,P})^\zeta\\ \nonumber
   N_{pp} &\propto&(N_{p,T}\,N_{p.P})^\zeta
\end{eqnarray}
where $N_{n,T}$, $N_{n,P}$ are the
numbers of neutrons in target and projectile,
and $N_{p,T}$, $N_{p,P}$ are the numbers of protons in target and projectile.
In intermediate energy heavy ion collisions, we find numerically that $\zeta$
has values roughly in the range between 2/3 and 1.

Inserting this {\em Ansatz} into our equation for the average cross section
yields the result displayed in Figure 2.  The difference in the systems with
different isospin is small (only approximately 0.5\% for $\zeta$=1), but is
clearly visible.  Since the isospin-averaged cross section is slightly
bigger for the $^{58}$Ni system, this results in a higher kinetic pressure,
therefore more repulsion, and thus a lower balance energy.

The isospin composition of the projectile and target also has an effect
on the Pauli exclusion principle for the final scattering states of the
nucleons.  If there are more neutrons than protons, then these neutrons
necessarily occupy are larger fraction of the available phase space for
the scattered nucleons, resulting in a smaller number of Pauli-allowed
$nn$ collisions than $pp$ ones.  However, for the considerations of the
isospin dependence of the average in-medium nucleon-nucleon cross
section, the interplay between Pauli-principle and isospin content of the
nuclei has only a small effect.  To convince ourselves of this, we note
that the Boltzmann collision integral contains the factor
\begin{equation}
      \Pi_{ab} = [1 - f_a(\vec r,\vec p)] \cdot [1 - f_b(\vec r,\vec p)]
\end{equation}
where $a$ and $b$ are the isospin indices of the two nucleons that take
part in the scattering event.  The factor $\Pi_{ab}$ represents the correction
of the free nucleon-nucleon cross section due to the Pauli exclusion principle
for the final scattering states.  If we average over all possible two-body
scattering events, we find for beam energies larger than the Fermi energy:
\begin{equation}
      \langle\Pi_{pn}\rangle \approx {\textstyle\frac{1}{2}}
      (\langle\Pi_{pp}\rangle + \langle\Pi_{nn}\rangle)
\end{equation}
The leading correction term to this approximation enters as $\delta^2$.
Thus the differential effect of the Pauli-exclusion principle on the
isospin dependence of the average in-medium cross section is only a small
correction to the determination of $\overline{\sigma}$.

\section{Numerical Results}

We preface this section by briefly outlining our procedure to determine
the balance energies:
We begin by calculating the proton freeze-out phase-space distribution at
reduced impact parameters $\hat{b}=b/b_{max}=0.275, 0.375, 0.475$ and $0.575$,
beam energies $50, 60, 70, 80, 90$ and $100$~MeV/A, using
Equations\,\ref{equation:limeanfield} and \ref{equation:sobmean}, and various
in-medium corrections $\alpha$, Eq.\ \ref{equation:corr}
This is followed by calculating the flow for
each permutation, and then the balance energy for each $\hat{b}$, using
a $\chi^2$ minimization procedure.

In Fig.\ 3, we show the experimental results of Pak {\it et al.} \cite{pak97}
as the shaded rectangles.  The width of these rectangles represent the
width of the impact parameter bins used for the integrations of the
experimental data.  The height of the rectangles represent the error bars
in the balance energy (standard deviation of the mean).  The darker shaded
rectangles represent the experimental results for the $^{58}$Fe + $^{58}$Fe
system, and the lighter gray areas those for the $^{58}$Ni + $^{58}$Ni.
The thick horizontal lines through the middle of each rectangle represent
the quoted experimental values.  This experimental information is the
same in all three panels of Fig.\ 3, each time compared to a different
calculation.

Before we discuss the ingredients of each calculation, we should point out
that the experimentally found balance energies in $^{58}$Fe + $^{58}$Fe 
system are always higher than those for the $^{58}$Ni + $^{58}$Ni system.
This is what we would expect from our elementary considerations of the
effects of the Coulomb interaction and the two-body collisions, but contrary
to what we expect from the mean field alone.  We will return to this point 
later.

The theoretical results for the $^{58}$Fe + $^{58}$Fe system are 
indicated by the open plot symbols, and those for $^{58}$Ni + $^{58}$Ni
by the filled ones.

In the left panel of Fig.\ 3, we show the result of a calculation with
the mean field interaction of B.A. Li \cite{pak97}, and with $\alpha=0$.
Pak {\it et al.} \cite{pak97} found that BUU under-predicted the balance
energies of $^{58}$Fe + $^{58}$Fe and $^{58}$Ni + $^{58}$Ni
collisions.  This is consistent with
previous work \cite{ogi90,kla93,west93} that
has shown the BUU model utilizing free-space scattering cross sections
consistently under-predicting the 
balance energies of
various systems.  However, the positive results of
Pak {\it et al.} was the correct reproduction of the differential
effect in the balance energies -- the difference between the
balance energies for the two systems has the right sign and approximately
the right magnitude.  This can be verified by examining the left panel of
Fig.\ 3.

In the central and right panels, we use the isospin dependent mean fields
of Li (center, \cite{li95,li96}, Eq.\ \ref{equation:limeanfield}) and
of Sobotka (right, \cite{sob94}, Eq.\ \ref{equation:sobmean}), combined
with the in-medium corrections to the elementary two-nucleon scattering cross 
sections.  For the central panel, we use $\alpha=-0.3$, and for the right
panel, we use $\alpha=-0.2$, respectively.

Let us first compare the results of the central and the left panel.  For
both calculations, we use identical isospin-dependent mean fields.  The only
difference is the change in the scattering in-medium correction.  We can
make two rather obvious observations:
\begin{enumerate}
\item The theoretical calculations are much closer to the data when using
the in-medium reduction of the scattering cross section -- for all impact
parameter intervals.  This observation
is consistent with previous results that did not look at isospin-dependent
effects \cite{ogi90,kla93,west93}.
\item Even though the balance energies for the iron system are still slightly
higher than for the nickel system, the magnitude of the splitting has been
reduced and is now at least a factor of 4 smaller than what is observed
in experiment.
\end{enumerate}

The same effects can be observed when using a different iso-spin dependent
mean field, compare the right panel of Fig.\ 3.  Here, the differences
between the theoretical balance energies are somewhat larger, but still
a factor of at least 2 smaller than those found in the data.

We mentioned above that the observed sign of the difference in balance
energies as a function of isospin is opposite to what one would expect
from the behavior of the isospin-dependence of the mean field.  This assertion
can be verified in our numerical calculations by switching the 
isospin-dependent term in the mean field on and off.  This comparison
has been performed in Fig.\ 4.  Here we use the same results as in the
right panel of Fig.\ 3 (Sobotka mean field, $\alpha=-0.3$, represented
by the diamonds in Fig.\ 4) and compare them
to a mean field without isospin dependence, but with the same value of the
nuclear compressibility (circles).  In particular for the larger impact
parameters, we can clearly see the expected effect:  The 
isospin-dependent part of the mean field actually {\em reduces} the
difference in the balance energies for the two systems.  This is consistent
with the results of our elementary considerations.  The three-times
larger isospin asymmetry in the iron system results in a stronger repulsion
due to the isospin dependent mean field and is thus pushing the balance
energy in the iron system more down than in the nickel system.

\section{Concluding Remarks}

We see an improvement in the performance of the BUU model's prediction of
balance energies as a function of impact parameter for collisions of $^{58}$Fe
+ $^{58}$Fe and $^{58}$Ni + $^{58}$Ni by including both an asymmetry energy
term in the mean field and in-medium reduction of the nucleon cross section.
We observe similar performance among the different formulations used for the
mean field: One formulation recently used by Bao-An Li \cite{li95,li96} 
and one due to Sobotka \cite{sob94}.  However, the mean field of Bao-An Li,
Equation\,\ref{equation:limeanfield}, requires $\alpha=-0.3$ in
Equation\,\ref{equation:corr}, whereas the Sobotka formulation, 
Equation\,\ref{equation:sobmean}, requires
$\alpha=-0.2$ for substantial improvement.
Efforts to distinguish between these two mean fields will
likely come from heavy ion calculations in conjunction with experiments near
the drip lines.  In the systems studied in the experiments by Pak {\it et al},
the range in isospin asymmetry was not sufficiently large to constrain
our parameter space any further.

We have shown the different contributions of the Coulomb force, the 
NN-scattering, and the isospin-dependent mean field on the isospin dependence
of the balance energy.  Both our elementary considerations and our numerical
results support the conclusion that the sign and the 
bulk of the isospin-difference in the
balance energies is caused by the difference in the Coulomb interaction and
the isospin dependence in the effective nucleon-nucleon two-body scattering.

While we can understand the absolute magnitude of the balance energies and
the sign of the isospin-dependent difference between the two systems,
our theoretical calculations do not yield the correct absolute magnitude
of the difference.  One can speculate on the origin of this important
disagreement.  A lack in our understanding of the isospin dependence of
the nuclear mean field or the isospin-dependent in-medium modification of
the two-body scattering cross section are the leading candidates.  However,
there is strong reason to suggest that the higher beam intensities of the
radioactive beam facilities currently in planning or under construction
will allow us to make progress in our understanding of this problem by
enabling us to explore a larger isospin asymmetry in heavy ion reactions.

\section*{Acknowledgment}

This work was supported in part by the National 
Science Foundation under grant PHY-9605207.

\clearpage

\clearpage

\begin{figure}[t]
\vspace*{12cm}

\includegraphics{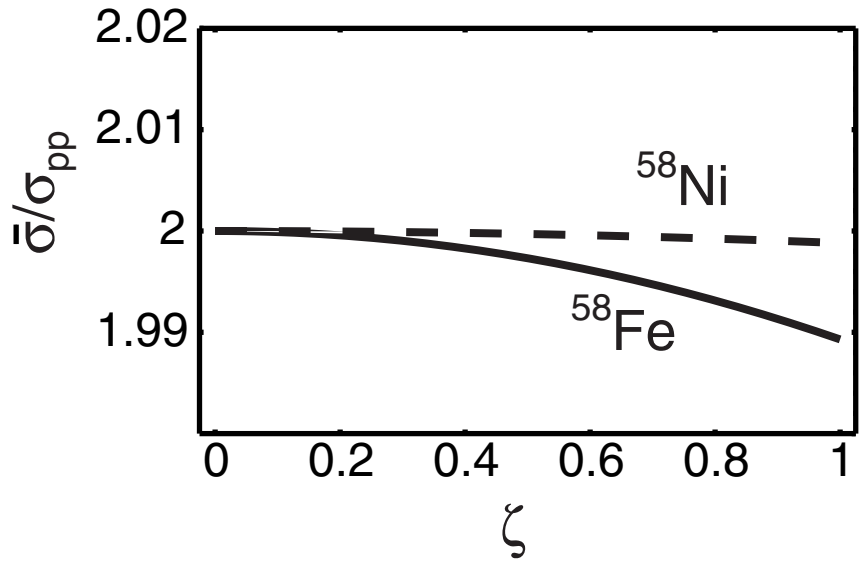}
\caption{Dependence of the average nucleon-nucleon cross section as a function
         of the parameter $\zeta$ for the two different isospin systems.}
\end{figure}

\clearpage

\begin{figure}
\vspace*{15cm}

\includegraphics{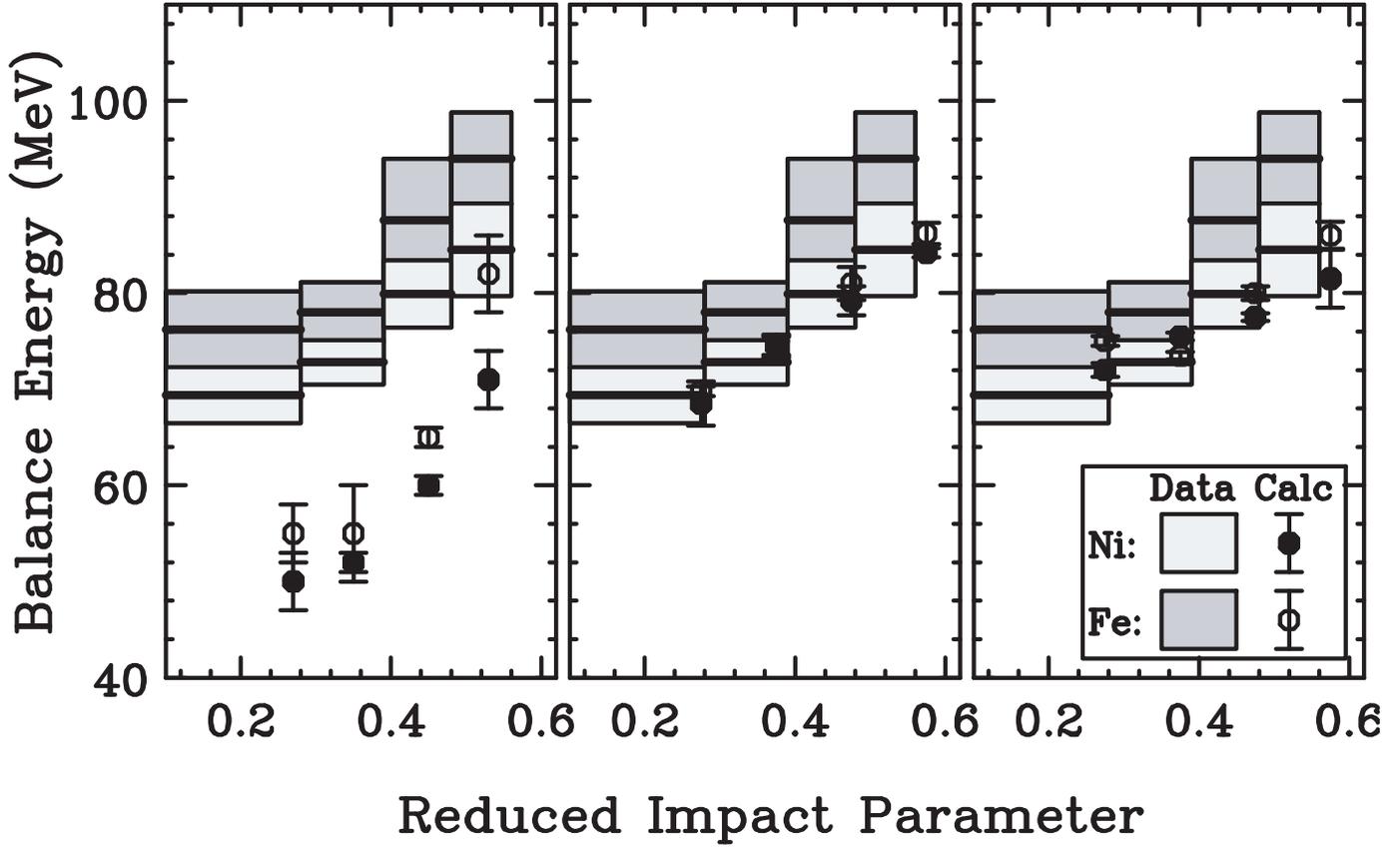}

\caption{Impact parameter dependence of the balance energy for the systems
      $^{58}$Ni + $^{58}$Ni (light shaded rectangles: data; 
      open circles: calculations) and
      $^{58}$Fe + $^{58}$Fe (dark shaded rectangles: data; 
      filled circles: calculations). Left panel: results of B.A. Li with
      free nucleon-nucleon cross sections; middle panel: mean field of B.A. Li
      and $\alpha$ = $-0.3$; right panel: mean field of Sobotka and
      \protect{$\alpha=-0.2$}.}
\end{figure}

\clearpage

\begin{figure}
\vspace*{15cm}

\includegraphics{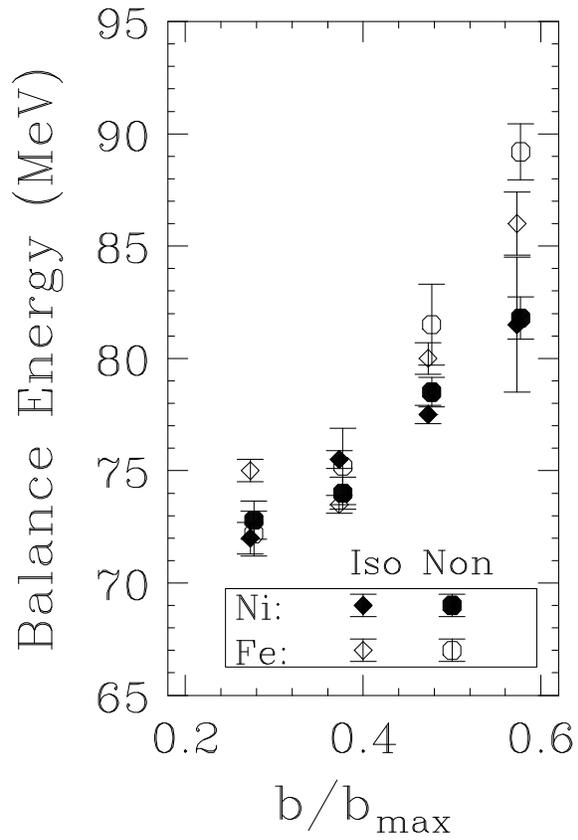}

\caption{Comparison of the effects of the isospin-dependent mean field
         (diamonds) relative to the corresponding mean field without
         isospin dependence (circles).}

\end{figure}

\end{document}